\documentclass{webofc}
\usepackage[varg]{txfonts}   
\usepackage{lineno}
\begin{document}

\newcommand{\hyt}{{\rm {}^{3}_{\Lambda}H}}
\newcommand{\sNN}{\sqrt{s_{\rm NN}}}
\title{Hypertriton Production in Au+Au
Collisions from STAR BES-II}

\author{\firstname{Yuanjing} \lastname{Ji}\inst{1}\fnsep\thanks{\email{yuanjingji@lbl.gov}}, for the STAR Collaboration} 

\institute{Lawrence Berkeley National Laboratory}
\abstract{
Hypernuclei are bound states of nuclei with one or more hyperons. Precise measurements of hypernuclei properties and their production yields in heavy-ion collisions are crucial for the understanding of their production mechanisms. The second phase of the Beam Energy Scan at RHIC (BES-II) offers us a great opportunity to investigate collision energy and system size dependence of hypernuclei production.
In these proceedings, we present new measurements on transverse momentum ($p_{T}$), rapidity (y), and centrality dependence of $\rm {}^{3}_{\Lambda}H$ production yields in Au+Au collisions from $\sqrt{s_{NN}}$ = 3 to 27 GeV. These results are compared with phenomenological model calculations, and physics implications on the hypernuclei production mechanism are also discussed.
}
\maketitle
\section{Introduction}
\label{intro}
\hspace{1.15em} Hypernuclei are bound nuclear systems of nucleons and hyperons. The presence of hyperons introduces an additional degree of freedom in baryon interaction: hyperon and nucleon ($Y$-$N$) interactions. Thus, hypernuclei are regarded as important probes to $Y$-$N$ interactions. The understanding of $Y$-$N$ interactions is important for constraining the strangeness degree of freedom of the Equation of State (EoS) in dense nuclear matter. In addition, the formation mechanisms of hypernuclei in heavy-ion collisions are of special interest in that the binding energies of hypernuclei are much smaller than the temperature of the system, e.g. the $\hyt$ binding energy $B_{\Lambda}\sim 100$ keV while the chemical freeze-out temperature $T_{ch}$ is of the order of $100$ MeV. The thermal model predicts that hypernuclei are abundantly produced in the low energy heavy-ion collisions above the $\Lambda$ production threshold since the baryon density increases as the collision energy decreases\cite{Andronic:2010qu}. A variety of observables are employed to investigate the hypernuclei production-related physics in heavy-ion experiments, e.g. the intrinsic properties (including lifetime, branching ratios, and binding energies, etc), the production yields, and the collectivity of the hypernuclei. 

\section{Analysis Details}
\label{sec-1}
\begin{figure}[htb]
\begin{minipage}{0.495\textwidth}
\centering
\vspace{0.1cm}
\includegraphics[width=0.81\linewidth]{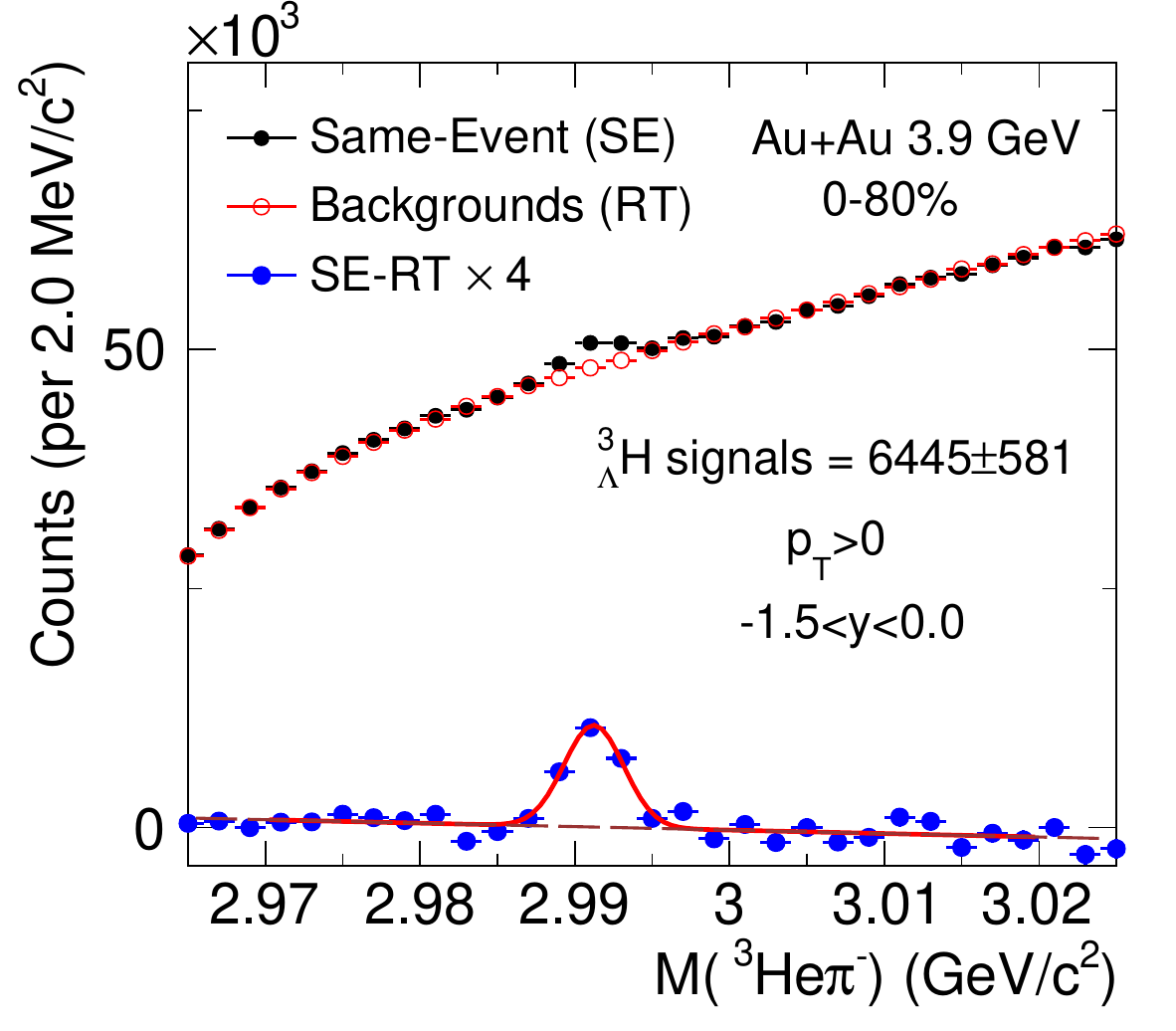}
\caption{The reconstructed $\hyt$ signal at $p_{T}>0$ and $-1.5<y<0$ in Au+Au collisions at $\sNN=3.9$\,GeV in 0-80\% centrality.}
\vspace{0.8cm}
\label{fig:signal}
\end{minipage}
\hspace{0.15cm}
\begin{minipage}{0.495\textwidth}
\includegraphics[width=0.765\linewidth]{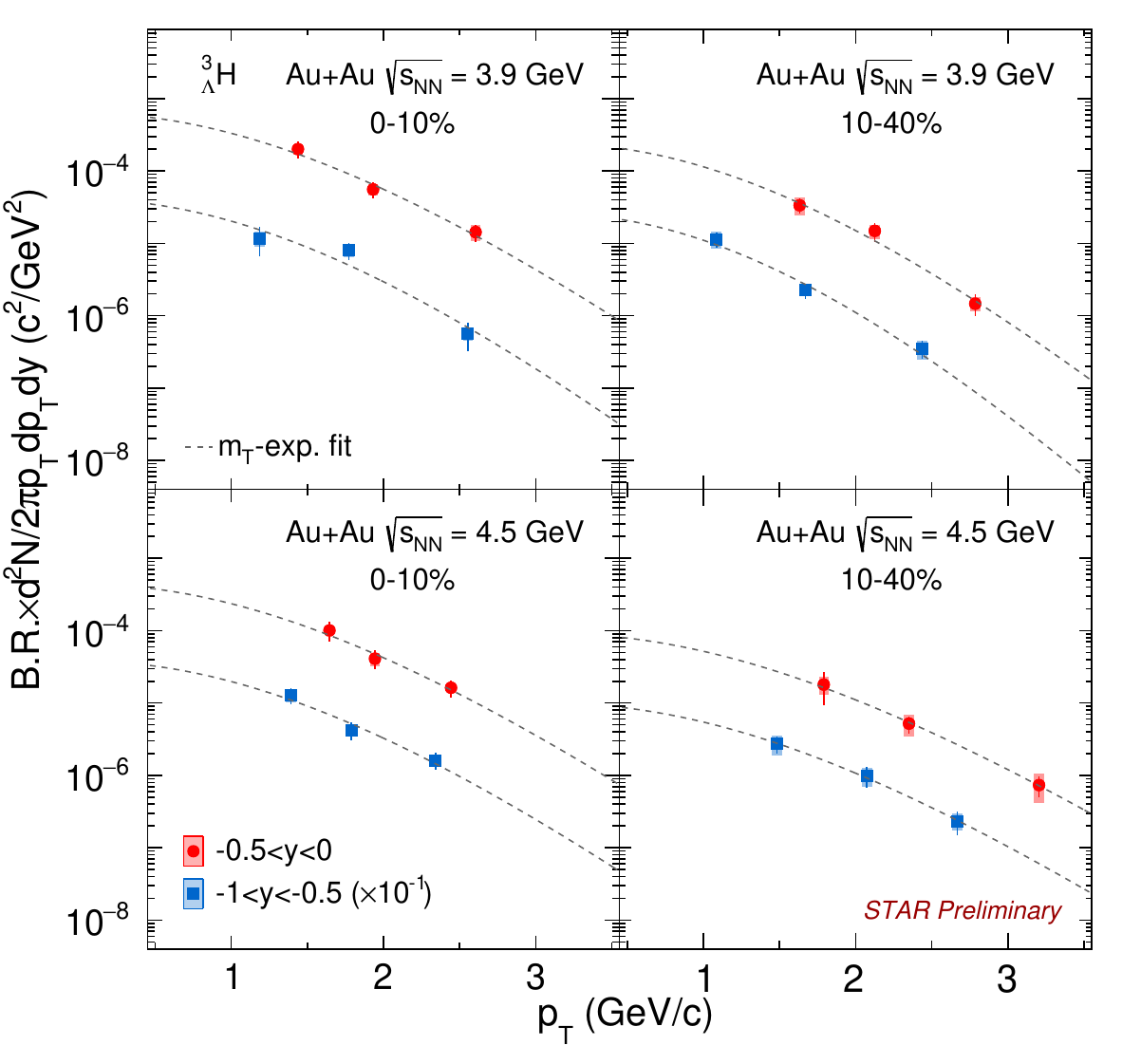}
\caption{The $\hyt$ $p_{\rm T}$ spectra in 0-10\% and 10-40\% centralites at $-0.5<y<0$ (red circles) and $-1<y<-0.5$ (blue squares) in Au+Au collisions at $\sNN=$ 3.9 and 4.5 GeV. The dashed lines are the fittings to the data.}
\label{fig:spectra} 
\end{minipage}
\end{figure}

\hspace{1.15em} The second phase of the Beam Energy Scan at RHIC (BES-II) collided gold nuclei (Au+Au) within a center-of-mass energy range from $\sqrt{s_{NN}}=3$ to 27 GeV. The program aims to systematically map the Quantum Chromodynamics (QCD) phase diagram, exploring the baryon chemical potential ($\mu_B$) within the range of $200 < \mu_B < 720$ MeV. In low energy collisions ($3\leq\sqrt{s_{NN}} \leq 7.7$ GeV), the collider ran under the fixed-target (FXT) mode to maintain high collision rates. The gold target is situated on the west side of the TPC detector which is the major tracking detector at the STAR. The TPC detector also serves as a particle identification detector by providing particle energy loss information (dE/dx). In these proceedings, the hypertriton $\hyt$ are reconstructed via the $\rm {}^{3}_{\Lambda}H \rightarrow{}^{3}He\pi^{-}$ and $\rm {}^{3}_{\Lambda}H \rightarrow dp\pi^{-}$ decay channels utilizing the KFParticle package \cite{Zyzak:2016exl,Ju:2023xvg}. Figure \ref{fig:signal} shows an example of the reconstructed $\hyt$ signals in 3.9 GeV Au+Au collisions at 0-80\% centrality. 

\vspace{-0.55cm}
\section{Results and Discussion}
\begin{figure}[htb]
\centering
\includegraphics[width=0.7\linewidth,clip]{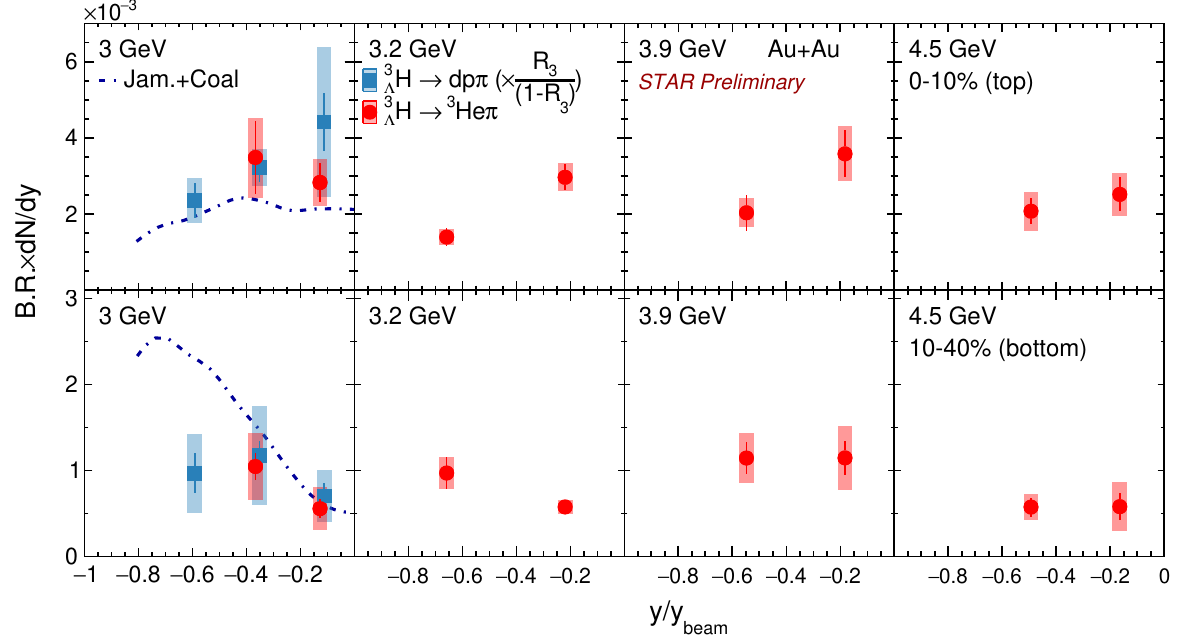}
\caption{The $\hyt$ dN/dy as a function of $y/y_{\rm beam}$ in Au+Au collisions at $\sNN=$ 3-4.5 GeV in 0-10\% and 10-40\% centralities. The $\hyt$ are reconstructed in $\hyt\rightarrow{}^{3}{\rm He}\pi^{-}$ (red circles) and $\hyt\rightarrow dp\pi^{-}$ (blue squares) channels. The dashed lines are the transport model JAM calculations with coalescence as an afterburner \cite{Liu:2019nii}.}
\label{fig:dndy}
\end{figure}

\begin{figure}[htb]
\begin{minipage}{0.495\textwidth}
\centering
\includegraphics[width=0.7125\linewidth]{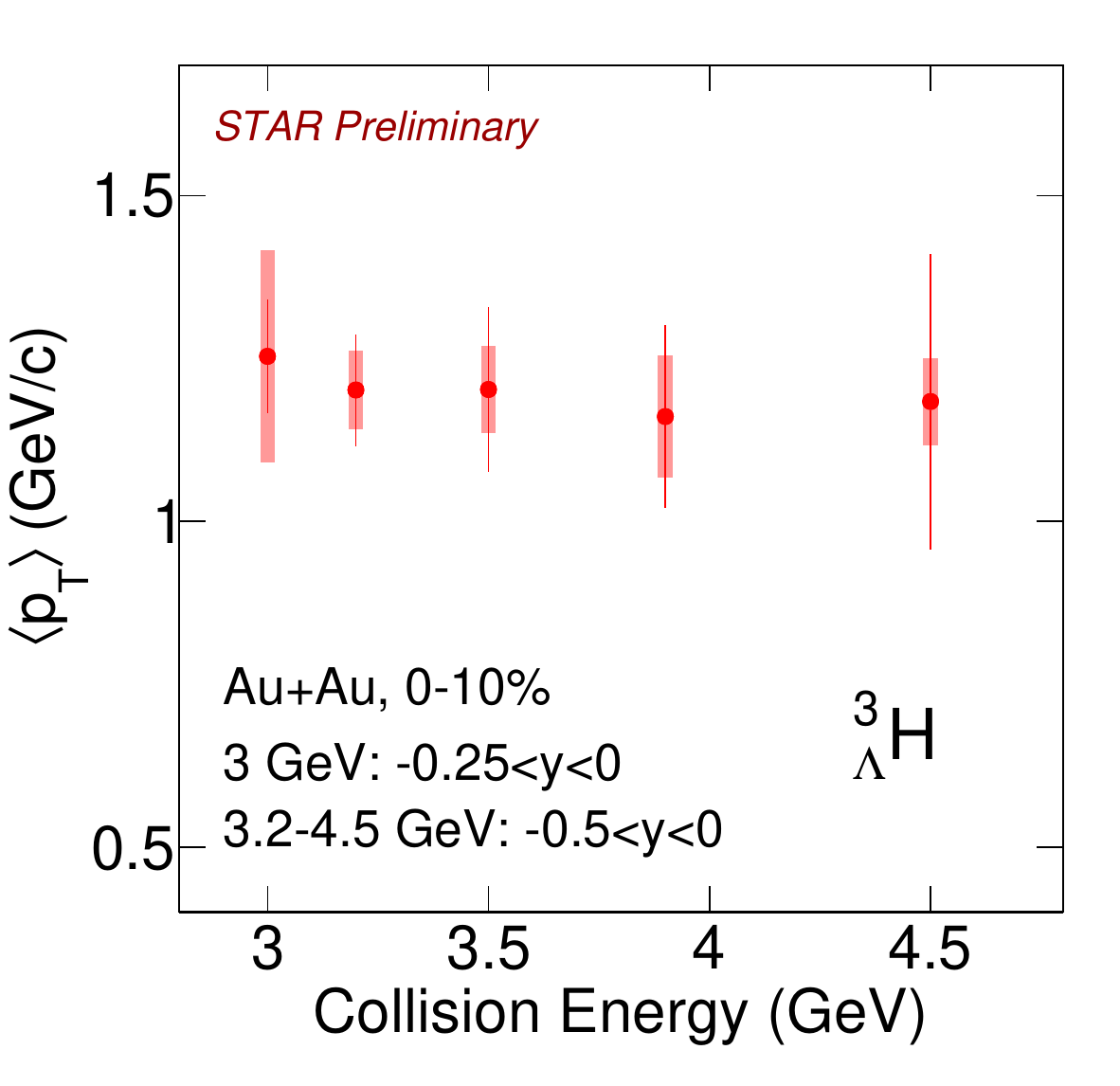}
\caption{The collision energy dependence of $\hyt$ $\langle p_{\rm T} \rangle$ at mid-rapidity in Au+Au collisions from $\sNN=$ 3 to 4.5 GeV in 0-10\% centralities.}
\label{fig:meanpT_energy}
\end{minipage}
\hspace{0.15cm}
\begin{minipage}{0.495\textwidth}
\vspace{-0.15cm}
\includegraphics[width=0.76\linewidth]{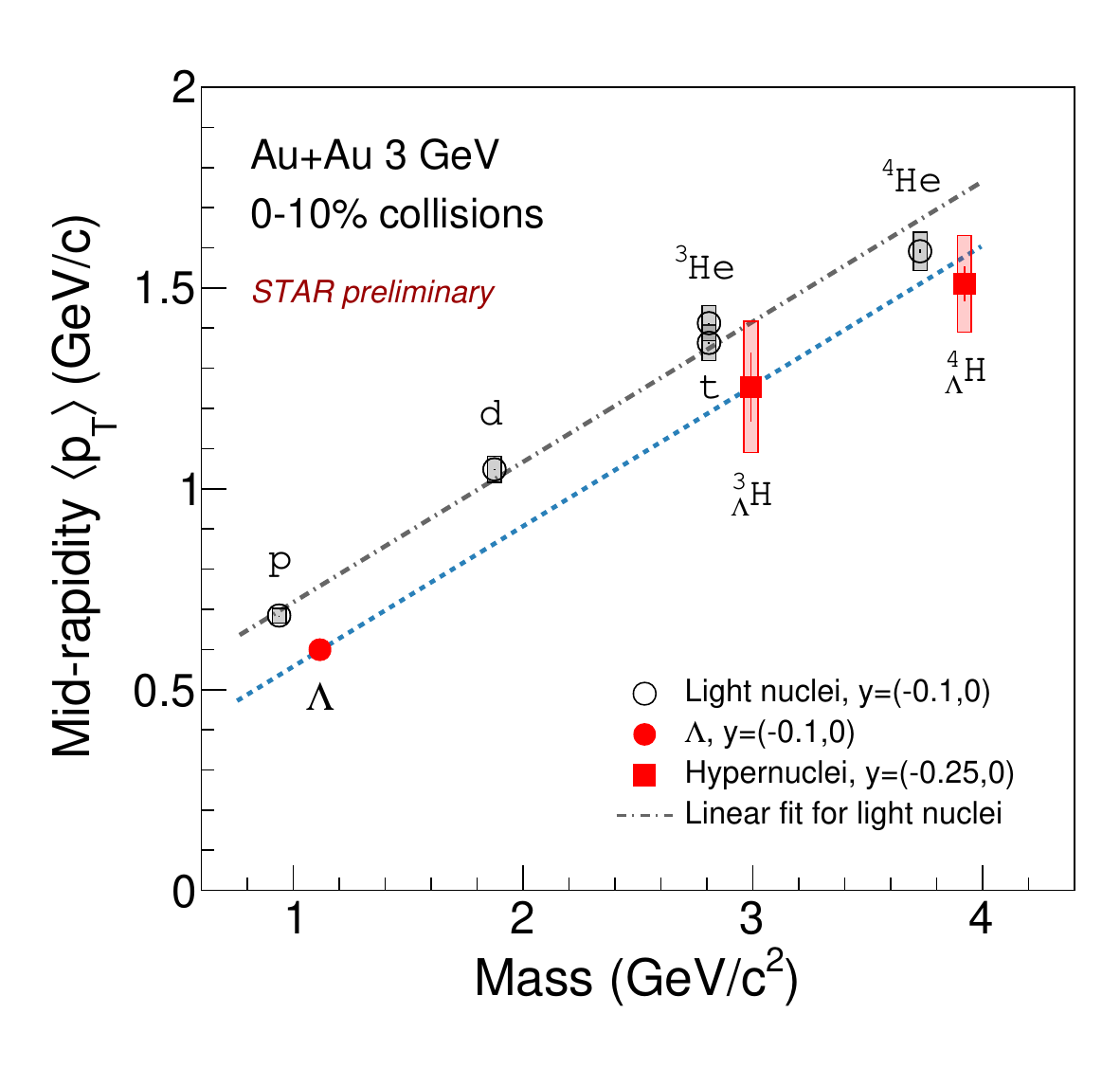}
\vspace{-0.15cm}
\caption{The $p$, $\Lambda$, light nuclei ($d$, $\rm {}^{3}He$, $\rm {}^{4}He$) and hypernuclei ($\hyt$, $\rm {}^{4}_{\Lambda}H$) $\langle p_{T}\rangle$ as a function of particle mass in Au+Au collisions at $\sNN=3$ GeV.}
\label{fig:meanpT_3GeV} 
\end{minipage}
\end{figure}

\hspace{1.15em} The hypertriton $p_{\rm T}$ spectra are measured in 0-10\% and 10-40\% centralities from $\sNN=$ 3-27 GeV. Figure \ref{fig:spectra} shows an example of the measured $\hyt$ $p_{\rm T}$ spectra in Au+Au collisions at $\sNN=$ 3.9 and 4.5 GeV. 
From the $p_{\rm T}$ spectra, the $\hyt$ mean $p_{\rm T}$ and dN/dy can be extrapolated. 
The $\rm {}^3_{\Lambda}H$ mean $p_{\rm T}$ and dN/dy are obtained from the $p_{\rm T}$ spectra using data in the measured $p_{\rm T}$ ranges and extrapolations assuming certain functional forms for the unmeasured $p_{\rm T}$ ranges  \cite{STAR:2021orx}.
Figure \ref{fig:dndy} summarizes the $\hyt$ dN/dy as a function of $y/y_{beam}$ from $\sNN=3-4.5$ GeV in Au+Au collisions in 0-10\% and 10-40\% centralities. 
Those comprehensive rapidity dependence measurments would enhance our understanding of the interplay between nuclear fragmentation and coalescence mechanism in hypercnulei production. 
In most central collisions, the transport model JAM with the instant coalescence of nucleons and hyperon as after burner \cite{Liu:2019nii,STAR:2022fnj} can qualitatively describe the rapidity dependence of $\hyt$ yields at $\sNN=$ 3 GeV in most central collisions, while it seems to fail to describe the trend in non-central collisions although the uncertainties in the data are still notable. Figure \ref{fig:meanpT_energy} shows the collision energy dependence of $\hyt$ $\langle p_{T} \rangle$ from $\sNN=$ 3-4.5 GeV. No significant energy dependence of $\hyt$ mean $p_{\rm T}$ is observed from $\sNN=$ 3-4.5 GeV. Figure \ref{fig:meanpT_3GeV} shows the $\hyt$ mean $p_{\rm T}$ in $0-10\%$ centralities in comparison with light nuclei and $\Lambda$ at $\sNN=$ 3 GeV. Both light nuclei and hypernuclei $\langle p_{\rm T}\rangle$ tend to follow mass scaling at $\sNN=$ 3 GeV within uncertainties. Similarly, the directed flow of light nuclei and hypernuclei are also observed to follow the mass scaling in 3 GeV Au+Au collisions within uncertainties \cite{STAR:2022fnj}. The mass scaling behavior of (hyper)nuclei $\langle p_{\rm T}\rangle$ and $v_{1}$ are qualitatively consistent with the expectations from the coalescence framework where the hypernuclei are formed via the coalescence of hyperons and nucleons.

\vspace{-0.5cm}

\begin{figure}[htb]
\begin{minipage}{0.495\textwidth}
\centering
\includegraphics[width=0.88\linewidth]{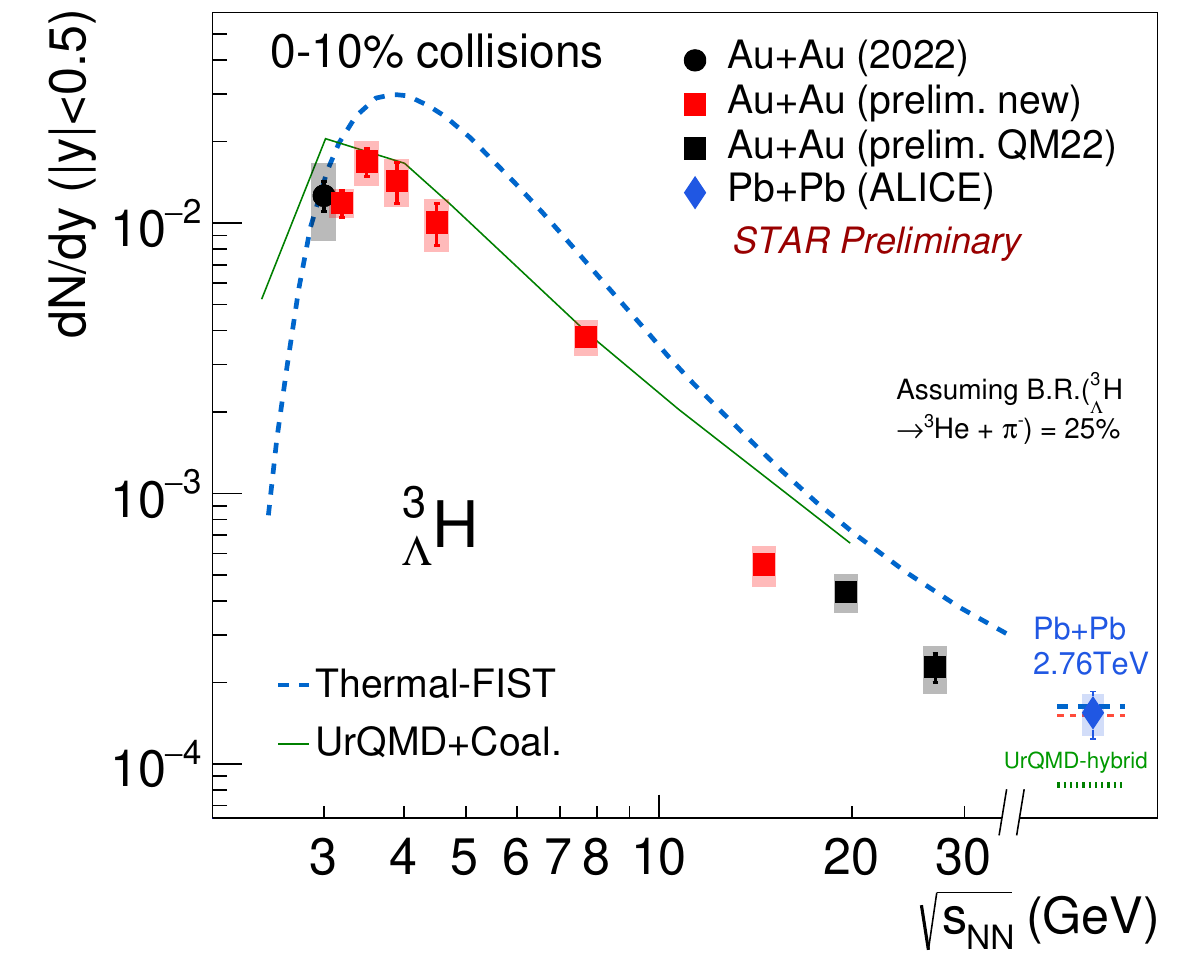}
\vspace{-0.2cm}
\caption{The energy dependence of $\hyt$ yields from $\sNN=$ 3-27 GeV in Au+Au collisions at 0-10\% centralities at $|y|<0.5$. The dashed line is from thermal model calculations \cite{Reichert:2022mek}. The solid line is from transport model calculations with coalescence as an afterburner \cite{Reichert:2022mek}.}
\label{fig:energy_yields}
\end{minipage}
\hspace{0.3cm}
\begin{minipage}{0.495\textwidth}
\vspace{0.15cm}
\includegraphics[width=0.88\linewidth]{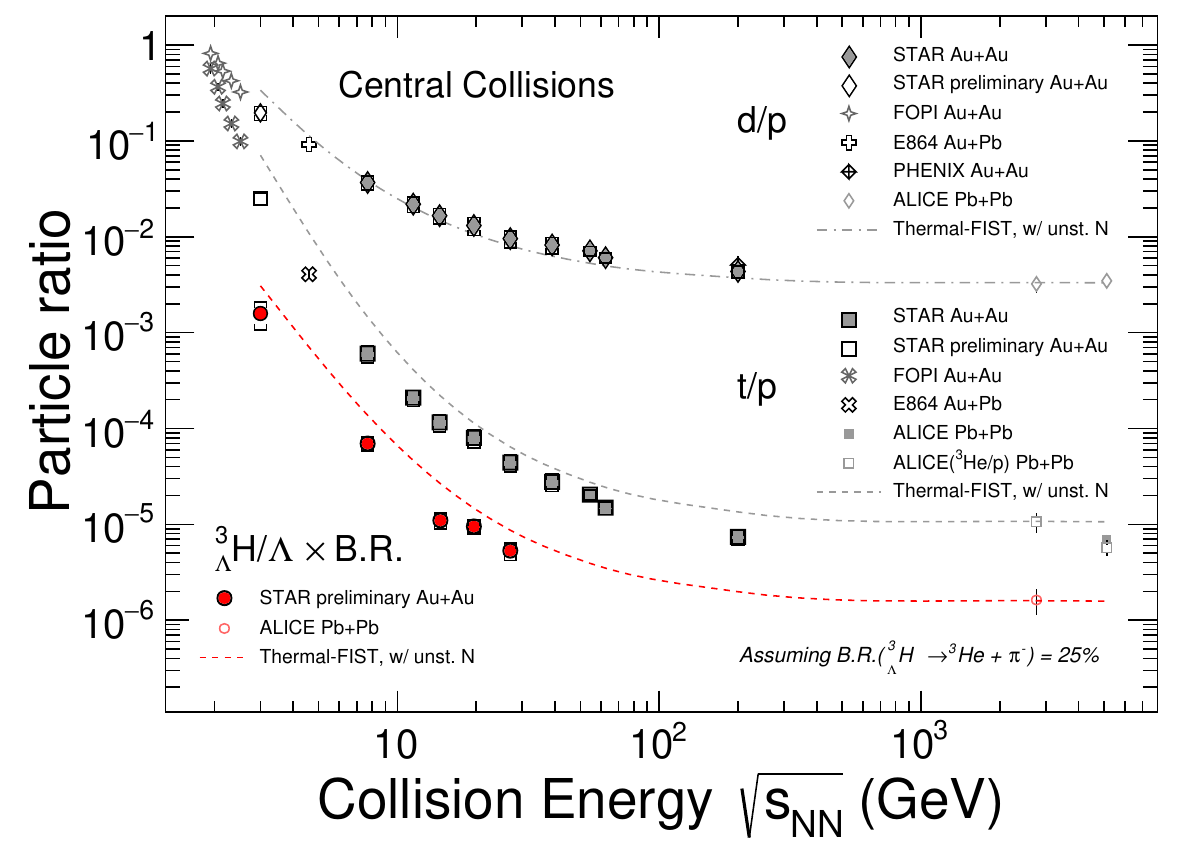}
\vspace{0.2cm}
\caption{The energy dependence of $d/p$, $t/p$ and $\hyt/\Lambda$ yield ratios from $\sNN=$ 3-27 GeV in Au+Au collisions at 0-10\% centralities at mid-rapidity. The dashed lines shown in the plot are from thermal model calculations \cite{Reichert:2022mek}.}
\vspace{0.3cm}
\label{fig:energy_ratio} 
\end{minipage}
\end{figure}

Figure \ref{fig:energy_yields} shows the energy dependence of $\hyt$ dN/dy at mid-rapidity 
from $\sqrt{s_{\rm NN}}=$ 3 to 27 GeV.
The $\hyt$ dN/dy in central Au+Au collisions increases as the collision energy decreases from 27 GeV to 4.5 GeV and then reaches the maximum at around $\sNN=$ 3-4 GeV. Two typical models from thermal \cite{Reichert:2022mek} and coalescence calculations (UrQMD+Coal.) \cite{Reichert:2022mek} are shown in Fig. \ref{fig:energy_yields}. The thermal model calculation assumes that the relative yield of particles composed of nucleons is determined by the entropy per baryon and the entropy is conserved after chemical freeze-out \cite{Andronic:2010qu,Andronic:2017pug}. The UrQMD+Coal. calculation firstly generates hadron phase space by hadronic transport model UrQMD. Then, based on the generated phase space, hyperons and nucleons would form into nuclei via instant coalescence if their relative momentum $\Delta P$ and coordinate $\Delta R$ are both less than model-dependent input thresholds \cite{Reichert:2022mek}. Both models can qualitatively describe the trends of the data while still having noticeable differences from the data central values. Figure \ref{fig:energy_ratio} shows the energy dependence of particle yield ratios for $d/p$, $t/p$, and $\hyt/\Lambda$. The dashed lines shown in the plot are from the thermal model calculations \cite{Reichert:2022mek}. The thermal model calculations can generally describe the $d/p$ ratio in the data while they are around 2 times higher than the data for both $t/p$ and $\hyt/\Lambda$ yield ratios. Those results indicate that hypertriton and triton yields might not reach equilibrium at hadron chemical freeze-out.

\section{Summary and Outlook}
\hspace{1.15em} In summary, we map $\hyt$ production yields in high baryon density regions in Au+Au collisions from $\sNN=$ 3 to 27 GeV. The hadronic transport model UrQMD with coalescence as afterburner and thermal model can qualitatively describe the measured energy dependence of $\hyt$ yields, while the thermal model is systematically higher than the data.
The current STAR mid-rapidity measurements on the hypernuclei production yields and collectivity favor the coalescence formation of hypernuclei in central collisions at mid-rapidity. 

The results presented in these proceedings utilize only a subset of the BES II datasets. During the RHIC run year 2021, STAR collected $2\times10^{9}$ events at 3 GeV which has $\sim 10$ times larger data size than that shown in these proceedings. The full BES II datasets and the new $\sNN=$ 200 GeV dataset (taken in 2023-2025) would enable precise measurements of light hypernuclei intrinsic properties. In addition, these datasets would extend the measurements of hypernuclei production to those hypernuclei with A$>$3. Combining all the STAR datasets, searching for the lightest double-$\Lambda$ hypernuclei is also a potential and ambitious project that might provide profound insights into hyperon-hyperon interactions.

\bibliography{ref.bib}
\end{document}